# A feasibility study of extruded plastic scintillator embedding WLS fiber for AMoRE-II muon veto

J.W. Seo[a, b], E.J. Jeon [a,b], W.T. Kim[a, b], Y. D. Kim[a,b,c], H. Y. Lee[b], J. Lee[b,*], M. H. Lee[a,b,**], P. B. Nyanda[a,b] and E. S. Yi[b,d]

[a]IBS School, University of Science and Technology (UST), Daejeon, 34113, Korea
[b]Center for Underground Physics, Institute for Basic Science (IBS), Daejeon, 34126, Korea
[c]Department of Physics and Astronomy, Sejong University, Seoul 05006, Korea
[d]Department of Astronomy, Space Science, and Geology, Chungnam National University, Daejeon, 34134, Korea

*Corresponding author (J. Lee):

E-mail address: jsahnlee@ibs.re.kr

**Corresponding author (M. H. Lee): Tel.: +82 42 878 8518; fax: +82 42 878 8509.

E-mail address: mhlee@ibs.re.kr

**Abstract**

AMoRE-II is the second phase of the Advanced Molybdenum-based Rare process Experiment aiming to search for the neutrino-less double beta decay of $^{100}$Mo isotopes using ~ 200 kg of molybdenum-containing cryogenic detectors. The AMoRE-II needs to keep the background level below $10^{-5}$ counts/keV/kg/year with various methods to maximize the sensitivity. One of the methods is to have the experiment be carried out deep underground free from the cosmic ray backgrounds. The AMoRE-II will run at Yemilab with ~ 1,000 m depth. However, even in such a deep underground environment, there are still survived cosmic muons, which can affect the measurement and should be excluded as much as possible. A muon veto detector is necessary to reject muon-induced particles coming to the inner detector where the molybdate cryogenic detectors are located. We have studied the possibility of using an extruded plastic scintillator and wavelength shifting fiber together with SiPM as a muon veto system. We found that the best configuration is two layers of plastic scintillators (PSs, 150 cm × 25 cm × 1.2 cm) with two WLS fibers per groove, which could separate radiogenic gammas well with muon detection efficiency above 99.4% along the length of the PS. Based on the expected flux from a prototype measurement at a 700 m deep underground, we found that the dead time of the muon veto system for AMoRE-II at the Yemilab with a 1 ms veto window is 0.6% of whole muon events.

Keywords: Neutrinoless double beta decay, SiPM, Muon veto detector, Underground experiment

**1. Introduction**

Above our heads, in the earth's atmosphere, many secondary particles are continuously produced by cosmic rays. These secondary particles can contaminate the low background experiments. For this reason, those experiments are being conducted deep underground, where these particles cannot penetrate. However, high-energy cosmic muons are deeply penetrating a kilometer of rocks. Muons can produce neutrons, producing gammas that mimic the signal close to the detector. Therefore, efficient identification of muons to remove associated signals is essential in the low background experiments, which is the primary goal of the muon veto detector.

Advanced Mo-based rare process experiment (AMoRE) is preparing the extended phase (AMoRE-II) after AMoRE-pilot and AMoRE-I for searching a neutrinos-less double beta decay in $^{100}$Mo-enriched scintillation crystals with cryogenic sensors [1]-[5]. The AMoRE-II detector comprises about 500 molybdate crystals (~100 kg of $^{100}$Mo) with cryogenic thermal sensors (metallic micro calorimeter, MMC) at tens of mK temperatures. We will install the cryogenic detector with a massive shield against external radiation in Yemilab at the Handuk mine under the Yemi mountain, Jeongseon, Korea. The depth of the Yemilab for ~1000 m rock overburden, assuming the standard rock density of 2.65 g/cm$^3$ [6], is estimated as ~2650 m w.e. AMoRE has put a lot of effort into removing various background sources, aiming for zero background in the region of interest (ROI), 3.024 ~ 3.044 MeV. The ROI is an energy range based on the detector's energy resolution goal of 10 keV (FWHM) centered at the Q-value (3.034 MeV) of $^{100}$Mo [3]. In particular, we reduced the internal background by developing a purification process for crystal raw materials [7]. In addition, the shield design using lead, polyethylene, and boric acid rubber is optimized to block radiation from outside. Despite the above various efforts, it is confirmed through a simulation study that the removal of muon-induced background is essential to achieving the zero-background goal [6]. We plan to install the muon veto detector on a larger scale for AMoRE-II compared to AMoRE-pilot and AMoRE-I. Several schematic views of the veto detector and their possible position can be found in Ref. [8].

Various muon veto detectors, including plastic scintillators, have been used in many other experiments. Scintillators are usually coupled to Photo Multiplier Tubes (PMTs) based on vacuum tube technology. PMT has been the most widely used device for detecting weak photon signals [9],[10]. However, silicon photomultiplier (SiPM) has become popular due to its compact size, insensitivity to magnetic fields, low cost, and low operating voltage, apart from the equivalent gain and quantum efficiency for PMT [11]. For scintillation detectors, the efficiency of collecting scintillation photons increases depending on the area of the photon sensor. However, the effective photo area of the largest single-chip (Multichannel) SiPM available as of today is still less than 36 mm$^2$ [12]. Using wavelength shifting (WLS) fiber in a large area plastic scintillator (PS), scintillation photons can be effectively collected, converted to higher wavelengths, and transferred to a small area SiPM, resulting in high muon detection efficiency [13],[14].

We will present the muon detector assembly method in section 2. Section 3 shows the surface lab measurements with different PS and WLS fibers combinations and an underground prototype measurement. Finally, the conclusion comes in Section 4.

**2. Muon detector assembly**

We decided to use the extruded PS for the AMoRE-II muon veto detector because it is simple, reliable, and efficient in detecting charged particles at a reasonable cost. But the extruded PS has a shorter attenuation length than the casting one [15].

A muon veto detector module comprises two extruded PS panels (CIMS-G2; CI Kogyo Ltd. [16]) with dimensions of 150 cm × 25 cm × 1.2 cm each. On the top side of the scintillator panel, there are 13 grooves in parallel along the length. The distance between the two adjacent grooves is 2 cm, and the depth is about 1.5 mm. The WLS fibers (Y-11(200) M; Kuraray Co., Ltd.) laid along the grooves have a diameter of 1 mm and are multi-cladding for higher light collection efficiency compared to the single cladding one [17]. The fibers are placed in the grooves without any optical grease and covered with a Vikuiti sheet using Scotch tape along the grooves (Figure 1). We polished both ends of each fiber with a sequence of 1500 and 2000 grit sandpapers. All fibers from thirteen grooves on a PS panel are bundled using an acrylic guide and optically coupled to a SiPM (S13360-6050CS; Hamamatsu Photonics K.K.) which has 14400 of 50 μm × 50 μm pixels in a 6 mm × 6 mm effective photosensitive area [18]. An optical grease (EJ-550; Eljen Technology) is used to couple fibers and SiPM (Figure 2). Each PS panel is wrapped with one layer of Tyvek (1082D; Dupont Co. [19]) as a reflector and a black polyurethane vinyl sheet [20] as a light shield.

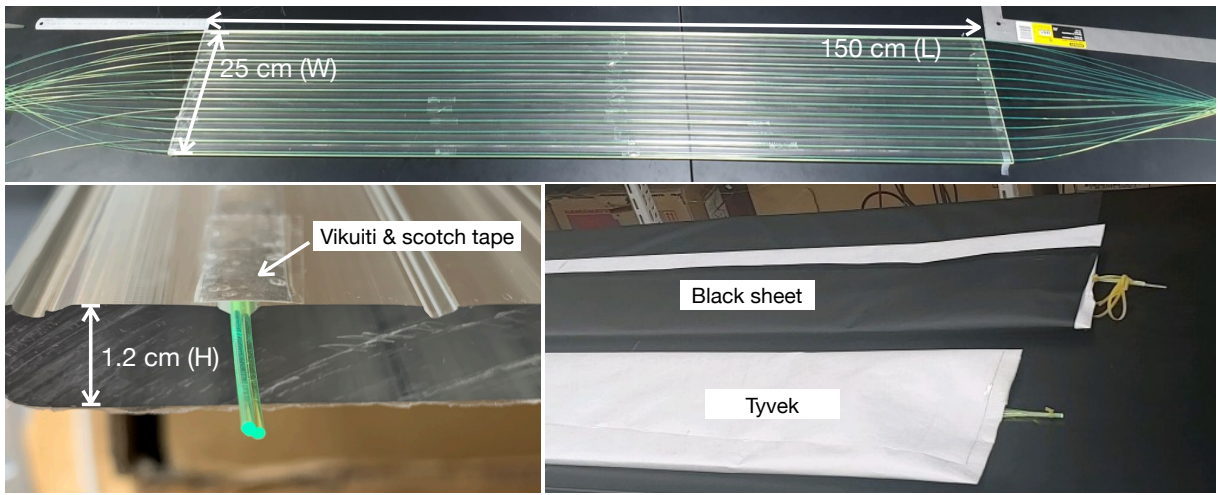

**Figure 1**: Pictures of a PS assembly with the WLS fibers. (Top) Thirteen grooves on an extruded PS have WLS fibers laid on. (Bottom left) Two WLS fibers in a groove are covered with a Vikuiti film using a scotch tape. (Bottom right) The assembled PS with WLS fibers is covered first by a Tyvek sheet and then by a black sheet.

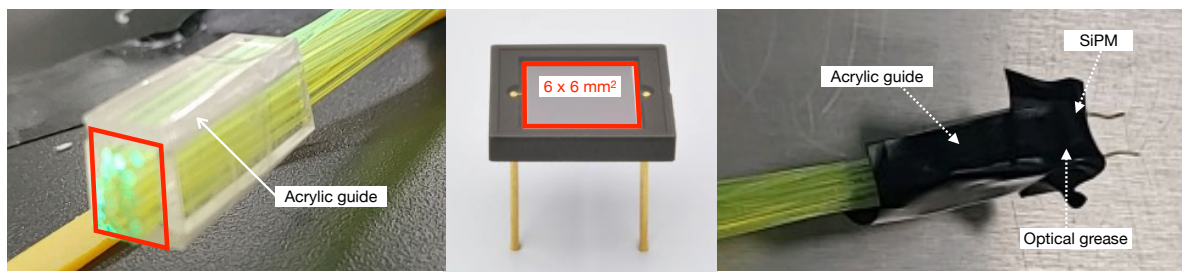

**Figure 2**: The pictures of fiber bundles in an acrylic guide, a SiPM, and a SiPM assembly.

As a minimum ionizing particle (MIP), the muon loses about 2 MeV per unit cm in a PS [21]. In the case of vertical incidence, about 2.4 MeV of energy is deposited in the PS with a thickness of 1.2 cm.

This energy is difficult to be distinguished from the energy left by the most energetic radiogenic gammas of 2.6 MeV. A muon detector module has two layers of PS to separate muons from the gammas by summing the deposited energies in both layers [22]. Since the Compton scattering by the gamma at one of the two layers reduces the probability of an event leaving high enough energies in both layers, we can significantly suppress the gamma background by requiring the coincidence of two scintillator layers.

Two front-end electronics boards are installed in a layer of PS and supply bias voltages to two SiPMs, read and amplify the signals from the two SiPMs. Typically, the breakdown voltage of SiPM(S13360-6050CS) is 53V, and then the overvoltage comes to be 2.5V. The front-end electronics board's signal is continuously digitized by a 12-bit flash ADC (NKFADC500; Notice Co.) with 500 mega sampling per second [23]. We generated a coincidence trigger when the four SiPMs signals of two muon counters are within the 64 ns time window. After calibrating the SiPMs to have the same gain, we set the trigger threshold at ~ 8 times single photoelectron (p.e.), low enough not to lose any muon signals.

## 3. Measurements

We took the muon data at the surface lab (IBS HQ, Daejeon) to test light yields for different PS and WLS fibers combinations and underground (Yangyang underground laboratory, Y2L) using the best configuration. In the following sections, we present all of the measurements in detail.

### 3.1 Light yield

The light yields are measured by changing the number of WLS fibers placed in a groove and the PS's and WLS fibers' lengths. To compare the different configurations, we measured the response of the detectors to the incident cosmic muons. The experimental setup is shown in Figure 3, where the light yield is measured for the "tested detector." A coincidence of two PS trigger counters, T1 and T2, is used to select the incident muon signals passing through the tested detector. Each PS trigger counter is a plastic scintillator with dimensions of 20 cm × 30 cm × 3 cm read out by a PMT (R329-02; Hamamatsu Photonics K.K.) via an acrylic light guide. A Landau distribution function fits the selected muon event data. The distribution's most probable value (MPV) is converted to the number of photoelectrons (NPEs) observed in SiPM by the light produced from an incident MIP in the scintillator.

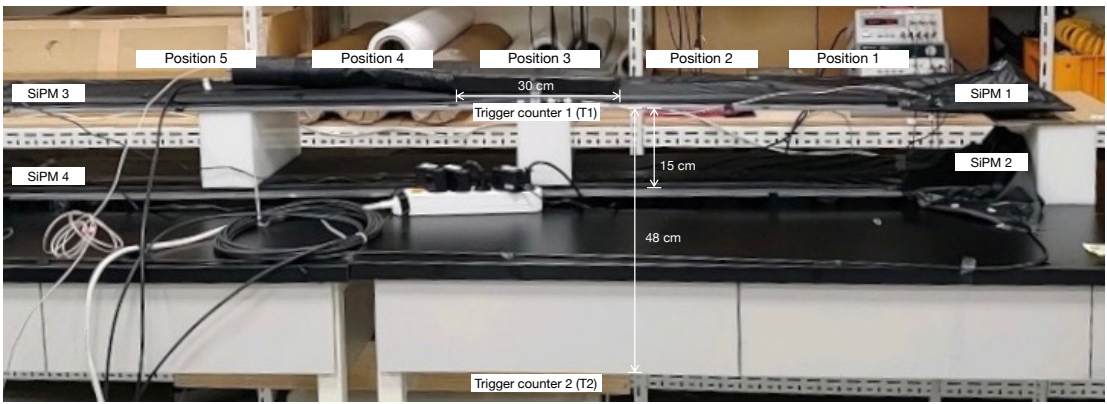

**Figure 3**: A test setup with two muon detectors and two PS trigger counters for the light yield and efficiency measurements at IBS HQ. We took the data moving the trigger counters to 5 different positions for an efficiency test.

The light yield depends on the number of fibers in a groove and the length of the fiber. Because the fiber has an attenuation length, the light yield decreases as the length of the fiber increases. The first configuration has a 150 cm long PS with one 240 cm long fiber per groove. Two 240 cm long fibers are in a 150 cm long PS groove in the second configuration. The third configuration has a 300 cm long PS with two 370 cm long fibers per groove. The light yields measured with the two PS trigger counters at position 3 (Figure 3) in the three different configurations are summarized in Table 1. The ADC value is converted to the number of photoelectrons after the single photoelectron signals of the SiPMs are measured. Distributions for the number of photoelectrons in the three configurations are shown in Figure 4. The peak position of the distribution represents the light yield of each detector configuration. We get more light yield with two fibers per groove than one. But the light yield in the longer PS and fibers configuration decreased. We took the following measurements on the surface and underground with the second configuration, which provides the most significant light yield.

**Table 1:** Light yields in three different configurations for the number of fibers per groove, fiber length, and PS length.

| Configuration number | The number of fibers per groove | Fiber length [cm] | PS length [cm] | Light yield [NPE/MIP] |
|---|---|---|---|---|
| 1 | 1 | 240 | 150 | 169.95 ± 0.16 |
| 2 | 2 | 240 | 150 | 306.32 ± 0.38 |
| 3 | 2 | 370 | 300 | 270.86 ± 0.27 |

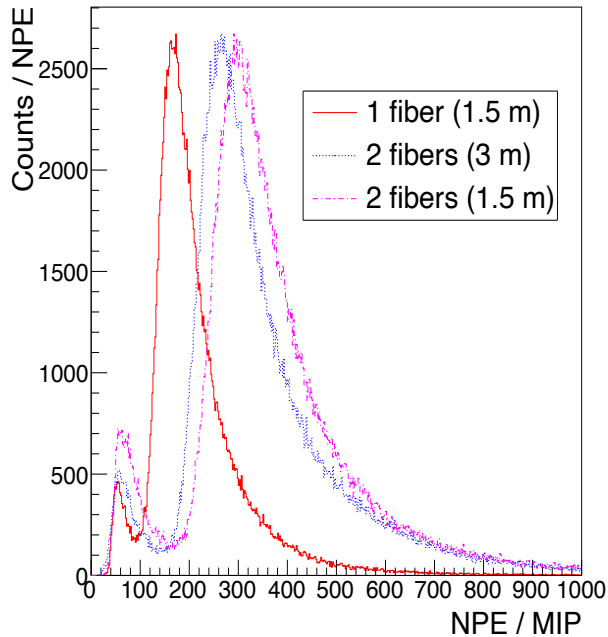

**Figure 4**: The number of photoelectrons distributions in three different configurations with different fiber lengths, different PS lengths, and number of fibers per groove as explained in Table 1.

### 3.2. Muon detection efficiency

The second configuration has the highest light yield among the three configurations. Muon detection efficiency is measured at five different locations in the longitudinal direction of the muon detector with the second configuration. A muon counter module comprises two PS panels. The distance between the two PS panels is 15 cm, as shown in Figure 3. We installed the two PS trigger counters, explained in the previous section, above and below the muon detector to count the number of muons passing through the muon detector under test. The trigger counters are placed in a way that the 20 cm sides are perpendicular to the longer sides of the extruded PS, and the 30 cm sides are parallel. The coincidence area was 20 cm × 30 cm, which was well within the muon detector under test. Data were collected from the muon detector using the coincidence of these two PS trigger counters. The number of passing through muon events was defined as the number of coincidence triggers. The position measurements were made by moving the PS trigger counters along the length of the muon detector by a 30 cm step. The center positions of the muon trigger counters were 15, 45, 75, 105, and 135 cm from one end of the muon counter.

The summed ADC values from four SiPMs correspond to the deposited energies of particles passing through the two PS layers in the muon detector. The events in the upper right area of Figure 5 (left), above 30,000 and 44,000 ADC values in T1 and T2, respectively, are regarded as muon signals. Figure 5 (right) shows the summation of the ADC values from all 4 SiPMs after the coincidence cut. We decided that the events above the 400 NPE are the detected muon signals in the muon detector.

We calculated the efficiency by counting the passing muon events and detected muon events. The obtained efficiencies for all five positions are shown in Table 2. It shows higher than 99.4% efficiency at any given position. It means that the possible light loss caused by the short attenuation length of the extruded PS can be complemented by using the WLS fibers.

**Table 2:** The efficiency values with statistical errors only for five positions were obtained by dividing the detected muon events into passing muon events.

| Position | Position 1 | Position 2 | Position 3 | Position 4 | Position 5 |
|---|---|---|---|---|---|
| Efficiency [%] | 99.45 ± 0.03 | 99.45 ± 0.03 | 99.40 ± 0.02 | 99.48 ± 0.02 | 99.40 ± 0.03 |

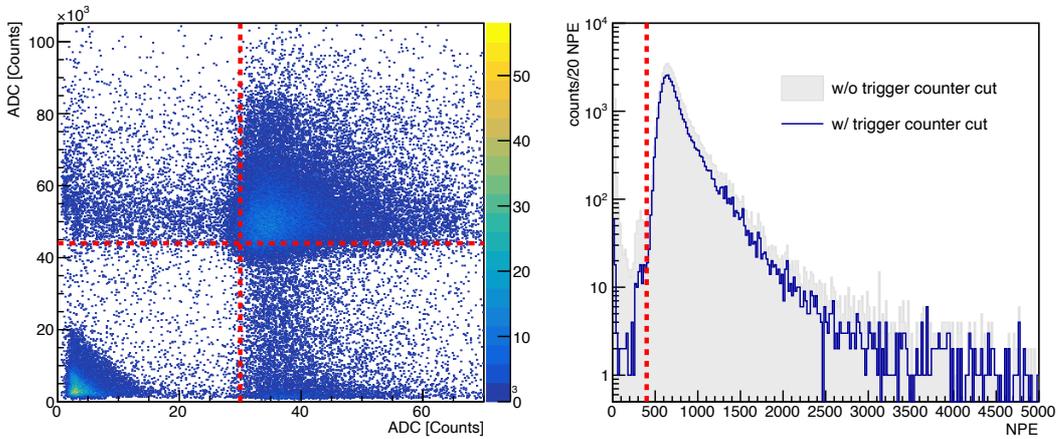

**Figure 5**: (left) A scatter plot of the two PS trigger counters' signals. (right) An NPE sum distribution from all 4 SiPMs in the muon detector module after a coincidence condition in the two PS trigger counters having ADC values above the (red) dotted lines on the left plot. The red dotted line is 400 NPE, events above which value are chosen as muons.

### 3.3. Underground measurement

The muon fluxes in the underground laboratories are significantly reduced [24]. The muon flux at the Y2L with 700 m of rock overburden was measured to be 328 ± 1 (stat.) ± 10 (syst.) muons/m$^2$/day by the COSINE-100 experiment [25].

After the measurement at the surface level, a prototype muon detector was installed in a room at the Y2L A5 tunnel where the COSINE-100 and AMoRE-I experiments are running [26]. In this measurement, two PSs with two fibers per groove are stacked upon each other. The DAQ system is the same as the one used in the surface measurement (Figure 6). All four SiPM signals from the two scintillators should exceed the trigger threshold at 40 ADC (~ 8 p.e.) within a 64 ns coincidence time window. After data taking, charge correction was performed before analyzing data.

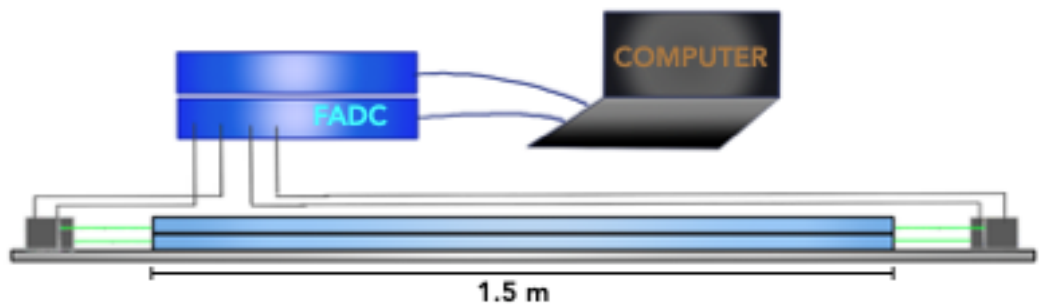

**Figure 6**: An experimental setup at Y2L. Two plastic scintillators are stacked up on each other. The signals from four SiPMs after shaping and amplification are sent to the FADC, which is connected to a DAQ notebook computer via a USB-3 connection.

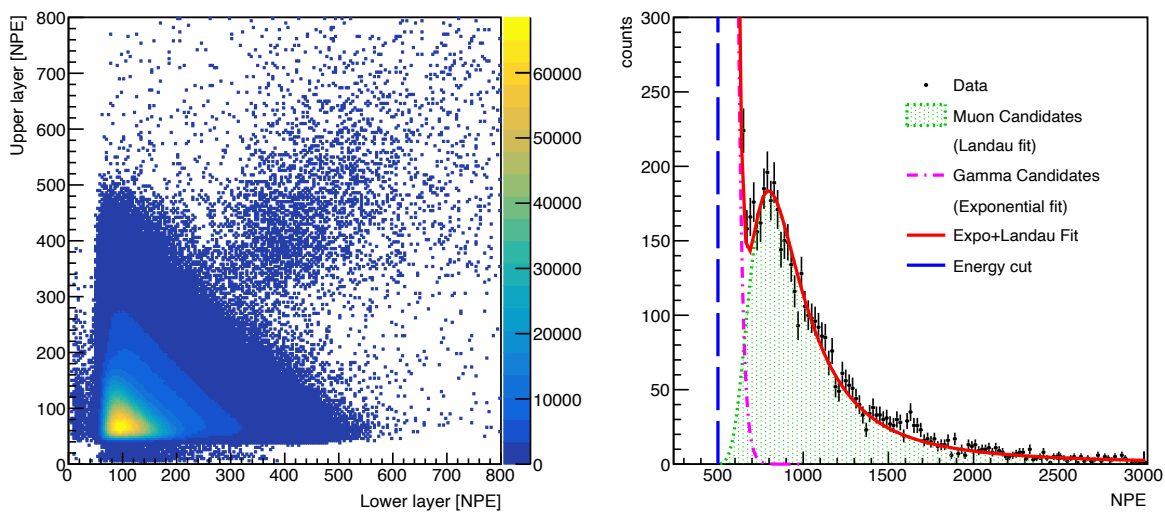

**Figure 7**: A scatter plot of the two scintillators' signals (left) and distribution of the NPE sum of the four SiPMs (right).

The NPE distribution of four SiPMs was analyzed to estimate the muon flux at Y2L. The exponentially decaying tail in Figure 7 (right) corresponds to the gammas and the peak to the muons. After fitting exponential and Landau functions to the distribution, we can obtain the number of muon-like and gamma-like events. We considered the signal sums above 500 NPE as muon candidate events for a safe vetoing. The obtained muon flux from the fitted Landau function is $457.6 \pm 6.6$ muons/m$^2$/day. The gamma background is exponentially distributed above the 500 NPE energy threshold. As a result, the muon veto event rate, including exponential gamma background from the 500 NPE cut, is 4421.3 events/m$^2$/day.

In our experiment setup, two PS counters stacked together with a minimal gap highly triggered by muons coming at higher zenith angles. An angular distribution of muons measured at Y2L was presented in [27]. The prototype detector is about 40 m apart from the COSINE-100 in the same tunnel, which is not enough to explain the flux difference. The flux difference also could be that the separation of muon signals from the background gammas is not enough with the prototype, as shown in Fig. 7. The final muon detector under construction has a thicker PS, 1.5 cm instead of 1.2 cm, to improve the separation.

AMoRE-II detector will be installed at the 1,000-m deep underground, ~1.5 times deeper than Y2L, of the Yemi mountain in Jeongseon, Korea. The integrated muon intensity at Yemilab can be normalized by the measured flux at the Y2L, assuming that their rock properties are equal for the two sites. Considering contour maps of Mt. Yemi and Yangyang areas, the muon intensity for both sites is 4.6 times different [6]. Thus, the expected muon flux of the Yemilab is 98.5 muons/m$^2$/day by scaling the Y2L measured flux with 4.6. Based on the expected muon flux on Yemilab, the muon event rate in the muon veto system of the AMoRE-II experiment (~135 m$^2$) is expected to be 62,208 muons/day. The veto event rate for the AMoRE-II muon veto system at the Yemilab is 548,640 events/day, assuming that the gamma background event rate is similar to Y2L. The dead time of the experiment depends on the veto window. This veto event rate corresponds to the dead time of 0.6% when we consider a 1 ms veto window.

## 4. Conclusion

AMoRE-II detector will be installed at the 1,000-m deep underground of the Yemi mountain in Jeongseon, Korea, different from the Y2L, where this prototype underground measurement was made. To reduce the muon background, we will install a muon veto system in the AMoRE-II. A prototype of the muon veto detector is designed and tested for different configurations in dimensions of PS and the number of fibers per groove using two SiPM readouts per one PS. The best configuration was two layers of PSs (150 cm × 25 cm × 1.2 cm) with two WLS fibers per groove, which could separate radiogenic gammas well with muon detection efficiency above 99.4% along the length of the PS. Based on the expected flux, we found that the dead time of the muon veto system for AMoRE-II with a 1 ms veto window is 0.6% of whole muon events.

**Acknowledgment**

This work was supported by the Institute for Basic Science (IBS), funded by the Ministry of Science and ICT, Korea (Grant id: IBS-R016-D1).